\documentclass{fizik}
\usepackage{hyperref}
\hypersetup{
colorlinks=true,
urlcolor=blue,
citecolor=blue}
\usepackage[all]{xy,xypic}
\usepackage{amsfonts,amssymb,amsmath,amsgen,amsopn,amsbsy,theorem,graphicx,epsfig}
\usepackage{eufrak,amscd,bezier,latexsym,mathrsfs,enumerate,multirow}
\usepackage[utf8]{inputenc}\usepackage[english]{babel}
\usepackage[dvipsnames]{xcolor}
\usepackage[pagewise]{lineno}

\yil{2023}
\vol{47}
\fpage{124}
\lpage{140}
\doi{10.55730/1300-0101.2742}

\title{Short-term optical variability of 4C 29.45}

\author[ÖZDÖNMEZ]{
\textbf{Aykut ÖZDÖNMEZ\thanks{aykut.ozdonmez@atauni.edu.tr}~\href{https://orcid.org/0000-0003-1399-5804}{}}\\
Department of Astronomy and Space Sciences, Faculty of Science, Atatürk University, Erzurum, Turkey
\\ [1.8em]

\rec{17.02.2023}
\acc{04.06.2023}
\finv{21.06.2023}
}

\newcommand{\bc}{\begin{center}}
\newcommand{\ec}{\end{center}}

\numberwithin{equation}{section}

\renewcommand{\phi}{\varphi}

\begin{document}

\maketitle

\begin{abstract}
We observed the flat-spectrum radio quasar 4C 29.45 in the BVRI optical bands for 39 nights from February 2022 to July 2022 with the T60 telescope at TÜBİTAK National Observatory (TUG) in Turkey. In this study, we aimed to study flux, color, and spectral variability on short timescales. The object was in an active (bright) phase with an average optical R-band brightness of 14.7 mag and was variable in the BVRI bands throughout the monitoring period. We analyzed the flux variability during our observation period, and the variability amplitudes in V, R, and I bands were determined to be $220\%$, $208\%$, and $209\%$, respectively.  Optical spectral energy distributions of 4C 29.45 were derived from the observational data of 33 nights, indicating spectral indices ranging from 1.032 to 1.573. We found modest correlations between optical light curves, and between R-band light curve and spectral indices, suggesting that the time lag ranges from several hours to days. {Investigation on relation between spectral and color indices versus R-band magnitude revealed achromatic trend during the bright phase of 4C 29.45 in the first half of 2022.} Our periodicity search suggested that the periodicity would be larger than 100 days, and no significant signal for periodicity was found for short timescales.
\end{abstract}

\section{Introduction}
A subclass of active galactic nuclei, blazars include two subtypes: flat-spectrum radio quasars (FSRQs) and BL Lacertae objects (BL Lacs).
The characteristic features of these objects are rapid variations in emissions across multiwavelengths, ranging from radio to gamma-ray, and significant variations in polarization \citep{1995PASP..107..803U, 1995ARA&A..33..163W}.
Unpredictable flares 
within the flux variability are detected in most of the blazars, and 
most blazars do not have reliable periodicity, with the exception of a few \citep
[e.g., OJ 287,][]{2008Natur.452..851V}. 
Flux variability of the blazars is  generally categorized considering diverse timescales \citep[e.g.,][]{2004MNRAS.348..831X, 2008AJ....135.1384G, 2022ApJ...933...42A}. 
The timescale from a few minutes to a day is referred to intraday variation (IDV) or microvariability. The amplitude of IDV is commonly a few of one-tenths of a magnitude. 
Short-term variation (STV) is generally referred for the variations over several months exceeding one magnitude. 
Lastly, long-term variation (LTV) is used for the variations with amplitude of a few magnitudes over a long timescale from a year to decades.
The exact causes of the flux variations and periodic behavior in blazars are still unknown. Possible explanations include shocks moving within the jet \citep
 {1996A&AS..120C.537M}, instability in the accretion disk \citep
 {1996ASPC..110...42W}, gravitational microlensing \citep
 {1987A&A...171...49S}, and variations in the Doppler factor due to the spiral movement of the emission plasma \citep{2017Natur.552..374R}.
Multiband observations are an indispensable tool in the study of blazars. These observations provide valuable information to get clues on the properties of the emission regions, including their size, structure, and dynamics as well as the cooling timescales of electrons in the relativistic jets.
Measuring flux and spectral variations by multiband observations play an important role in verifying theoretical models \citep
 [e.g.,][]{2003A&A...400..487C, 2008AJ....135.1384G, 2015MNRAS.450..541A}.

The variation of optical flux in blazars is often accompanied by color changes that provide useful insights into their emission mechanisms. In general, three different color trends are identified: bluer-when-brighter (BWB), redder-when-brighter (RWB), and achromatic.
A frequently observed trend in BL Lac objects is the BWB. Two possible explanations for this trend are the shock-in-jet model and Doppler factor variations  \citep[e.g.,][]
{1997A&A...327...61G, 2002A&A...390..407V, 2010MNRAS.404.1992R,2019MNRAS.488.4093A}. 
RWB is observed more frequently in FSRQs and is believed to be due to additional radiation from the accretion disk \citep[e.g.,][]{2006A&A...453..817V}. 
The spectral properties of the FSRQ depend on various factors, such as the frequency of the synchrotron peak, the optical wavelength interval, and the relative strength of the thermal blue emission compared to the emission from the jet \citep{2011A&A...528A..95G}. {The common explanation for achromatic behavior is often attributed to variations in the Doppler factor. These variations are likely best explained within the context of the geometric scenario \citep{2002A&A...390..407V}.}
Alternative trends have also been suggested \citep[e.g.,][]{2016MNRAS.458.1127G, 2017ApJ...844..107I, 2021A&A...645A.137A,
 2022MNRAS.510.1791N, 2000ApJS..127...11G, 2006MNRAS.366.1337S, 2007ApJ...670..968B,
 2009ApJS..185..511P}, including cycle/loop-like patterns where BWB is dominant during high states and RWB during low states, or a stable-when-brighter (SWB) trend where there is no significant color-magnitude correlation at any timescale.

The blazar 4C 29.45, also known as 1156+295 and Ton 599, is a highly variable FSRQ emitting multiwavelength radiation from radio to $\gamma$-ray
\citep{1983ApJ...274...62W, 1992ApJ...398..454W,
2006PASJ...58..797F, 2007A&A...469..899H,
2008A&A...489L..33S,2022ApJ...926..180H}. 
It has a core-jet structure and has been observed by various campaigns due to its high-energy emissions \citep[see][and the references therein]{1995ApJS..101..259T,2014MNRAS.445.1636R, 2017ApJ...846...98J,2022ApJ...926..180H}. 
From these observations, characteristic measurements of jets including
Doppler factor, Lorentz factor, and jet viewing angle were determined as $\delta=12\pm3$,  $\Gamma=10\pm3$, and  $\theta \leq 2^{\circ}.5$, respectively
\citep{2017ApJ...846...98J}. 
Polar clouds located at a few parsecs away from the central supermassive black hole are subjected to ionization by UV photons from synchrotron radiation coming of upstream regions within the jet \citep{2022ApJ...926..180H}.
The $\gamma$-ray variations were found to lead the optical variations by about --0.8 days with a correlation coefficient of 0.70  \citep
{2022ApJ...926..180H}. 
The blazar exhibits violent variability up to 6 mag in both the optical and infrared bands on diverse timescales \citep{1983ApJ...274...62W, 1996ASPC..110..170B, 1996ASPC..110...30N,
2000ApJS..127...11G, 2006PASJ...58..797F}. Possible periods  were found to be 1.58 or 3.55 years \citep{2006PASJ...58..797F}. It is currently in its second year of enhanced optical activity with frequent flares, including a recent flare that brightened by 1.5 magnitudes in a week and reached R = 13.4 mag
on June 19, 2022 (UTC) \citep[][and references therein]{2022ATel15441....1D}.

In this study, we present the results of our optical observations of the blazar 4C 29.45 in the first quarter of 2022, including the examination of its latest flare. 
{To gain insight into the radiative mechanisms of the blazar, we had carried out quasisimultaneous observations in the BVRI bands using the 0.6m telescope at TÜBİTAK National Observatory (TUG) in Turkey.} The focus of our analysis was on the multiband STV, the correlation between the optical bands, periodicity of the STV, and spectral behavior. Understanding the complex nature of blazars requires multiwavelength observations on various timescales, and our observations contribute to this broad mission.

\section{Materials and methods}
\subsection{Observations and data reduction}
The optical properties of 4C 29.45 were examined by quasisimultaneous observations in the BVRI bands conducted from February to July 2022, spanning a total of 146 days and utilizing 39 nights of data collection (see Table \ref{tab:obs_log}). 
A 60-cm robotic telescope (T60) at the TUG was used for the observations. It is equipped with a 2k $\times$ 2k CCD camera and 12 standard filters\footnote{https://tug.tubitak.gov.tr/en/teleskoplar/t60-0}, while the robotic telescope is capable of taking daily photometric images of the target field with a few sets of BVRI frames as long as the weather permits.
The exposure times of the frames ranged from 30 to 120 s depending on the band and source brightness. {TUG uses TALON software pipeline for the T60 data correction procedure including bias/dark correction and flat-fielding. Thus, we only performed cosmic-ray removal on the corrected CCD data, and then the aperture photometry to extract the instrumental magnitudes of the blazar and standard stars in the field by our algorithms using Python and Photutils package \citep{2021zndo....596036B}.} Various concentric aperture radii were tested, ranging from 1.0 to 3.0 times the full width at half-maximum (FWHM) of the standard stars, while the sky annulus was set to 5 and 7 $\times$ FWHM to subtract the background.
The instrumental magnitudes of the sources were extracted based on the observation chart presented in \citet{1985AJ.....90.1184S}.

To perform further analyses, we employed instrumental magnitudes that were obtained from an aperture radius of 1.4 times FWHM. This aperture radius was selected based on the criterion of providing the optimal signal-to-noise ratio and the lowest standard deviation of the differential instrumental magnitudes of the standard stars. The instrumental magnitudes of the target object were calibrated using standard stars 13 and 14 in the chart of \citet{1985AJ.....90.1184S}. These stars were chosen due to both their brightness and their position closest to the blazar.

\subsection{Variability amplitude}
To evaluate the variability of the light curves (LCs), the variability amplitude was used as a metric. The variability amplitude is defined as follows:

\begin{eqnarray}
A = \sqrt {(A_{max} - A_{min})^{2} - 2\sigma^{2}},
\end{eqnarray}

where $A_{max}$ and $A_{min}$ denote the maximum and minimum magnitudes during the observation period, and $\sigma$ is the mean error. This definition is taken from \citet{1996A&A...305...42H}.

\subsection{Discrete correlation function}
In order to detect any correlation between the emission in the optical bands and to determine the potential lags, the discrete correlation function (DCF) was utilized \citep{1988ApJ...333..646E}. The DCF is a widely employed statistical technique for investigating correlations between two unevenly and/or irregularly sampled time-series data, particularly in the study of active galactic nuclei. It has been used extensively in this field, as evidenced by several studies \citep[][and references therein]{2007A&A...469..899H, 2015MNRAS.450..541A, 2017ApJ...841..123P, 2021MNRAS.504.1427A}.

The DCF between two discrete data sets (ai, bj) can be computed using the unbinned DCF (UDCF) defined as follows:
\begin{eqnarray}
UDCF_{ij}(\tau) = \frac{(a_i-\bar{a}) (b_j-\bar{b})}{\sqrt{(\sigma_{a^2}-e_{a^2})(\sigma_{b^2}-e_{b^2})}},
\end{eqnarray}
where $\bar{a}$ and $\bar{b}$ represent the mean values of the two time-series data sets, $\sigma_{a,b}$ and $e_{a,b}$ are their standard deviations and errors, respectively. $\Delta t_{ij}=(t_{bj}-t_{ai})$ denotes the time delay between the two data points. To obtain the DCF, the UDCF values are averaged over the interval $\tau - \frac{\Delta\tau}{2} \leq \tau_{ij} \leq \tau + \frac{\Delta\tau}{2}$ for {each time lag $\tau$} as described in the study by \citet{1988ApJ...333..646E}).

\begin{eqnarray}
    DCF(\tau) = \frac{\sum_{k=1}^{m} \text{UDCF}_k}{M}.
\end{eqnarray}
The quantity "M" represents the count of pairwise time lag values situated within the specified $\tau$ interval.

The DCF transforms into a discrete autocorrelation function (DACF) when a single time series is used. It measures the similarity between a data point and a lagged version of itself. By computing the DACF for different lags, it is possible to identify the most prominent periodic patterns in the data and determine the periodicity of the variability. The method for detecting periodicity in blazar light curves has been used commonly \citep[e.g.,][]{2004A&A...424..497V, 2021MNRAS.504.1427A, 2022ApJ...933...42A}.

\subsection{Lomb--Scargle Periodogram}
The Lomb--Scargle periodogram (LSP) is a commonly used statistical tool for detecting periodic signals in time series data. It was first introduced by Lomb and Scargle in the 1970s and 1980s \citep{1976Ap&SS..39..447L, 1982ApJ...263..835S}. This method involves fitting sinusoidal waves to the data at each frequency using a form of least-square fitting. The LSP is particularly effective for data with unevenly spaced observations, making it a more appropriate choice compared to the conventional discrete Fourier transform. The resulting periodogram features, particularly peaks, may indicate the presence of periodic signals in light curves. For the purpose of detecting periodicity of variability, the Lomb--Scargle periodograms package of Astropy\footnote{https://docs.astropy.org/en/stable/timeseries/lombscargle.html} was used in this study.

\subsection{Spectral and color relations}
\label{sec:2_5}
Investigating the variation of spectral indices obtained from the spectral energy distribution (SED) and the color trend obtained from color-magnitude (CM) diagrams can be achieved by monitoring the object in the optical BVRI bands. By resolving the flux and color variability, we gain insight into the radiation mechanisms of the blazar emission and get opportunity to explore various variability scenarios.

The color trend can be obtained by analyzing the correlation between the color indices and brightness. {This was commonly achieved by conducting a linear regression analysis in the literature
\citep[e.g.,][]{2016MNRAS.458.1127G, 2017ApJ...844..107I, 2021A&A...645A.137A,
 2022MNRAS.510.1791N}}, where the optical color indices (CI) were plotted against the R-band magnitude, using the equation $CI = m R + c$. Here, $m$ represents the slope and $c$ represents the constant of the regression fit. {Here, we considered the color configurations of $B-I$, $B-R$, $R-I$, and $V-I$.} The strength of this correlation was measured by computing the Spearman correlation coefficient ($r$) and null hypothesis probability ($p$).

To understand the variability in the spectral properties of 4C 29.45, we collected multiband optical spectral energy distributions (SEDs) from data of nearly simultaneous observations in all four bands. In order to analyze the results accurately, the calibrated BVRI fluxes had to be cleaned up for galactic extinction ($A_\lambda$) using the NASA/IPAC Extragalactic Database\footnote{https://ned.ipac.caltech.edu}. The values of interstellar extinction in the BVRI bands, as provided by the database, were low and listed as $A_B = 0.072$ mag, $A_V = 0.053$ mag, $A_R = 0.042$ mag, and $A_I = 0.030$ mag. The corrected magnitudes were derived by subtracting the interstellar extinction from the calibrated BVRI magnitudes and then converting them to corresponding fluxes based on the zero-points outlined in \citet{1998A&A...333..231B}. The optical SEDs were found to follow a simple power law of the form $F_{\nu} \propto \nu^{-\alpha}$, where $\alpha$ is the optical spectral index. The spectral indices for each night were calculated by fitting a linear relationship of the form $log(F_{\nu}) = -\alpha~log(\nu)+ C$.

\section{Results and discussion}
\subsection{Flux variability}

\begin{figure}
\centering
\includegraphics[width=0.9\linewidth,clip=true]{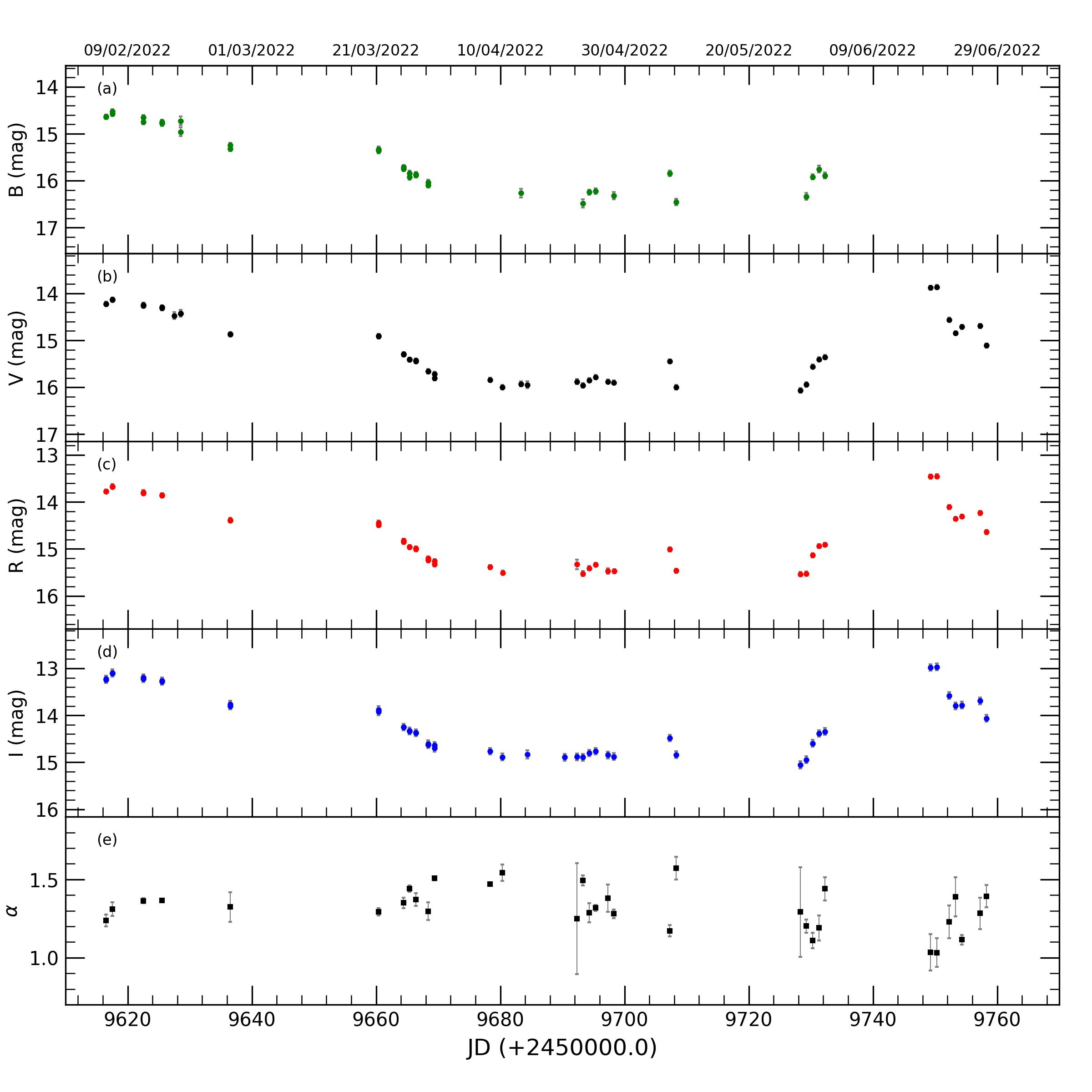}
\caption{\label{fig:STV} Short-term light curves of 4C 29.45 in $B$, $V$, $R$, and $I$ bands are represented in (a), (b), (c), and (d) panels, respectively. The variation of spectral indices $\alpha$ based on time is given in (e) panel.}
\end{figure}

The simultaneous observations in multioptical bands for blazars provide valuable insights into emission regions, particle acceleration mechanisms, and radiation processes. We observed the blazar 4C 29.45 long enough
to examine the short-term variability (STV) from 2022 Feb to 2022 July. The STV light curves of the source in optical BVRI bands during this period are shown in Figure \ref{fig:STV}, {which includes $\alpha$-time variation as well.} As reported in \citet{2022ATel15441....1D}, the blazar 4C 29.45 exhibited a flaring period in June, which resulted in a 1.5 mag increase in brightness within a week, reaching R=13.4 mag on June 19, 2022.
We also spotted peak to ending stage of the latest
flare in June in BVRI bands as well. 
Although our observations did not cover the full extent of the flare, we were able to detect the peak and ending stages of the flare in BVRI bands. The brightest magnitude was recorded in the R band, reaching a value of $13.451\pm0.041$ mag on $\sim$ MJD 59750.3 (19 June 2022). The flare ended around $\sim$ MJD 59758.5 (27 June) with $R=14.639\pm0.42$ mag and the fading rate was estimated to be 0.14 mag/day, resulting in a dimming of 1.188 mag in 8 days. During the brightest state (flaring period as well), the optical magnitudes in the V, R, and I bands were $13.864\pm0.043$, $13.451\pm0.041$, and $12.967\pm0.071$ mag, respectively, while  the observations in the B band were unavailable.  
At its faintest state on June 28-29, 2022, the B, V, R, and I band magnitudes were measured as $16.480\pm0.089$, $16.066\pm0.046$, $15.534\pm0.046$, and $15.054\pm0.074$ mag, respectively.
The results of the STV variability are listed in Table \ref{tab:STV_res}.
The  average $VRI$ magnitudes of 4C 29.45 for our entire monitoring period were
estimated as $\Bar{V}=15.147\pm0.046$, $\Bar{R}=14.705\pm0.044$, and $\Bar{I}=14.161\pm0.072$ mag.
The percentage of variability amplitudes of the STV were calculated as
follows: 220.155\%, 208.243\%, and 208.692\% for V, R, I, respectively.
Because there is no B band data during flare, we could not estimate the brightest
magnitude, average magnitude, and variability amplitude for B bands.

{\citet{2006PASJ...58..797F} compiled the optical data of the
blazar from the available literature and their own observations performed during the period of 1997 to 2002, thus obtained historical long-term light curve covering 1907--2002. 
Long-term light curve showed that the optical variations of the blazar 4C 29.45 are 
$\Delta B=5.55$ mag (18.28--12.73 mag), 
$\Delta V = 4.53$ mag (17.38--12.85), 
$\Delta R = 5.80$ mag (18.04-12.24), and 
$\Delta I = 5.34$ mag (17.03-11.69). 
\citet{2022ApJ...926..180H} also monitored the source in optical R band between December, 2014 and June, 2019 obtaining brightness ranging from 18.4 to 13.4 mag ($\Delta R=5$ mag). Between 1907 and 2019, the historical faintest and brightest magnitudes of 4C 29.45 have been measured as $R=18.4$ and $R=12.24$ mag ($\Delta R=6$ mag), respectively. Thus, our observations in the first half of the year 2022 clearly included high (bright) phase, and the brightest stage of the source was $\sim$ 1 mag fainter than historical flare peak during our observation period.
Consequently, 4C 29.45 was in bright phase and showed optical flux variations mildly ($\Delta R=2.1$ mag) on the monthly intervals throughout the observation period.}

\subsection{Spectral and color variability}\label{sec:spectral}
\begin{figure}
\centering
\includegraphics[width=0.75\linewidth,clip=true]{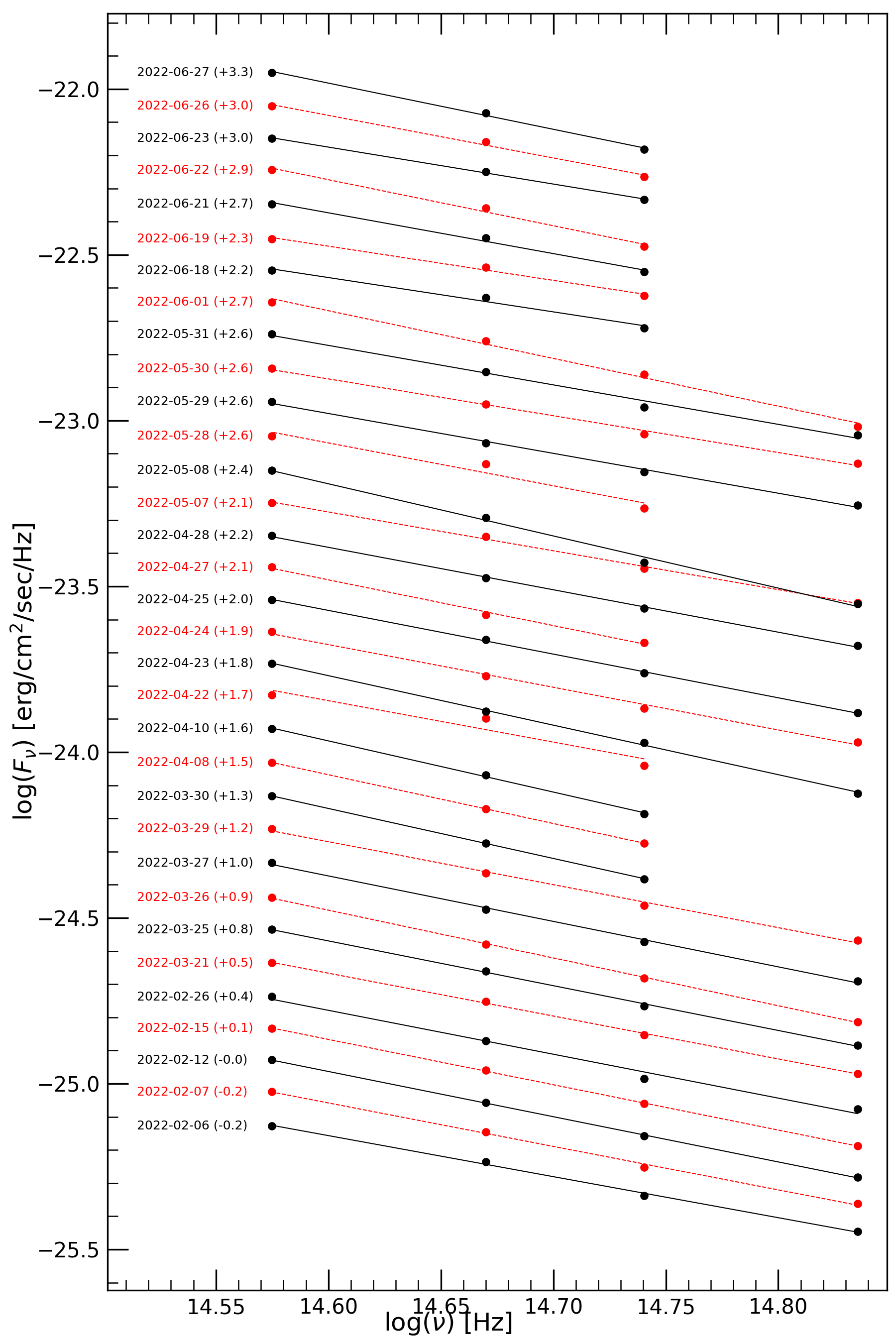}
\caption{\label{fig:sed_fig} Nightly optical SEDs of 4C 29.45 for $B$, $V$, $R$, and $I$ bands.}
\end{figure}

Studying spectral variation is crucial for understanding the source of variability in blazars. FSRQs are typically low-energy peaked blazars with a synchrotron peak at IR-optical frequencies. 
Thus, the study of optical variability can provide insights into different theoretical models. Variations in the spectra of emitting electrons are the main cause of SED modifications, resulting from differences in the physical properties of relativistic jets.
To study the daily multiband optical SEDs of the blazar 4C 29.45, simultaneous observations in VRI bands were extracted. There were 33 nights with simultaneous VRI datasets and 21 nights with BVRI fluxes. The optical SEDs were derived for the 33 nights as shown in Figure \ref{fig:sed_fig}. Linear regression was applied to each SED to obtain the spectral indices, which are listed in Table \ref{tab:sed}. {It should be noted that the small degrees of freedom due to the small number of data points during the fitting process increased uncertainties of the correlation as expected. However, there is a statistical difference in optical fluxes based on frequencies only at three dates with p-values larger than 0.05.}
The findings of the linear regression analysis indicate that the spectral index varied between 1.032 and 1.573, with an average spectral index of $1.313 \pm0.070$. Additionally, the standard deviation was determined to be 0.13. The calculated spectral indices based on time are displayed in Figure \ref{fig:STV}e. {There is no significant trend in $\alpha$ versus time  diagram, and visibly there is no relevancy with the optical light curves.}

\begin{figure}
\centering
\includegraphics[width=0.65\linewidth,clip=true]{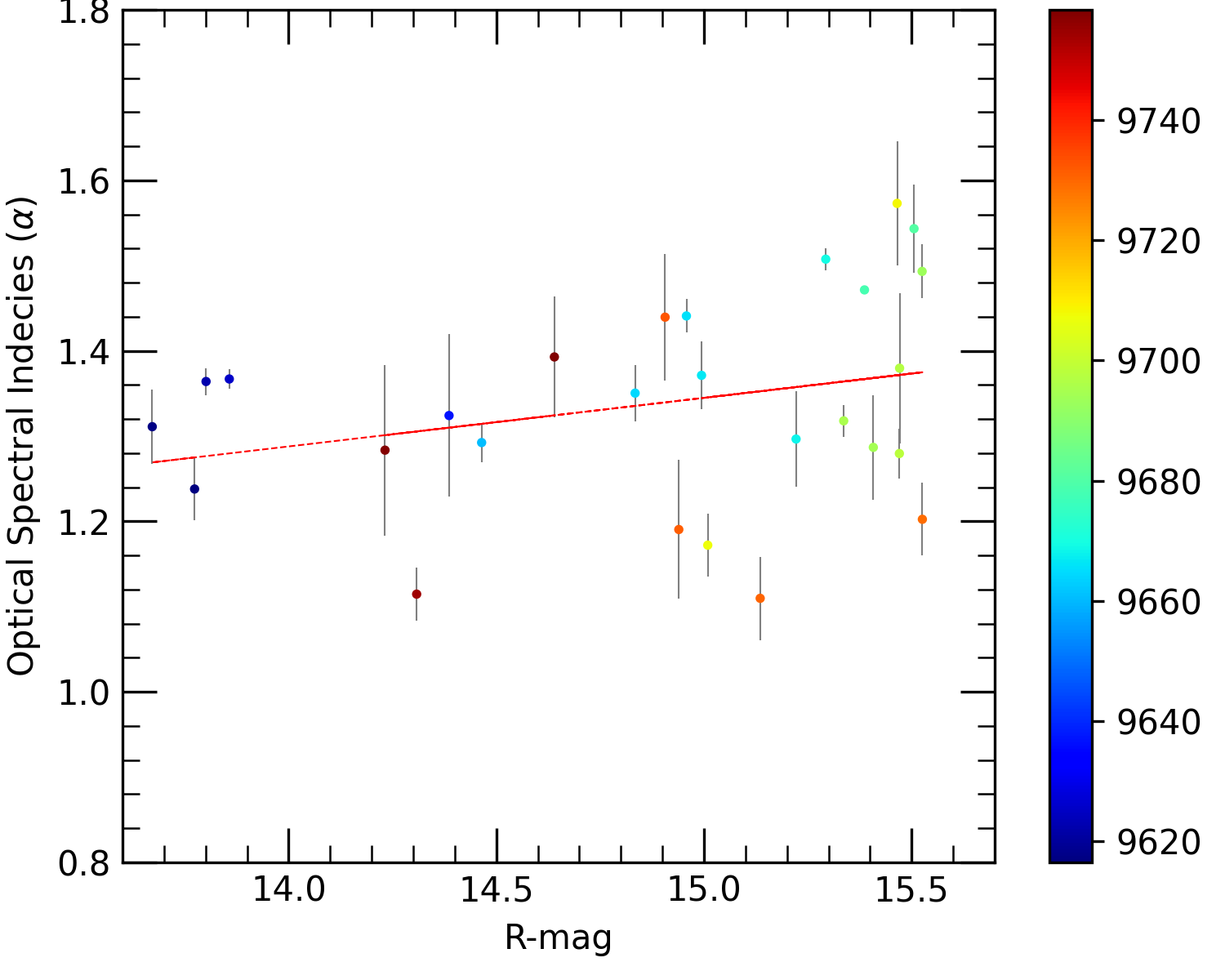}
\caption{Variation of nightly optical spectral indices of the blazar 4C 29.45 with respect to R-mag during the entire monitoring period. Color of points represent the JD of the observation as shown in color-bar for scale of JD-245000.} 
\label{fig:alpha}
\end{figure}

In order to understand the spectral variations in blazar 4C 29.45, the spectral index distribution ($\alpha$) with respect to the R-band magnitude is shown in Figure \ref{fig:alpha}. Subsequently, a linear regression analysis was performed to examine the correlation between $\alpha$ and the optical light-curve in R-band. The results of the regression analysis yielded a slope of $0.0089\pm0.032$, a constant of $-0.005\pm0.47$, a Spearman correlation coefficient of $r=0.45$, and a null hypothesis probability of $p=0.0085$. 
These findings indicate a weak correlation ($r\lesssim 0.5$) between the spectral indices and R-band magnitude. The positive slope in the spectra suggests that the spectral shape of the object flattens as it gets brighter. 
{However, achromatic trend could be considered if the absolute value of the slope is less than the corresponding fitted uncertainty, i.e. very small slope. 
Therefore, a  blue-when-brighter (BWB) trend is nonsignificant, and achromatic trend could be adopted.
}

\begin{figure}
\centering
\includegraphics[width=0.85\linewidth,clip=true]{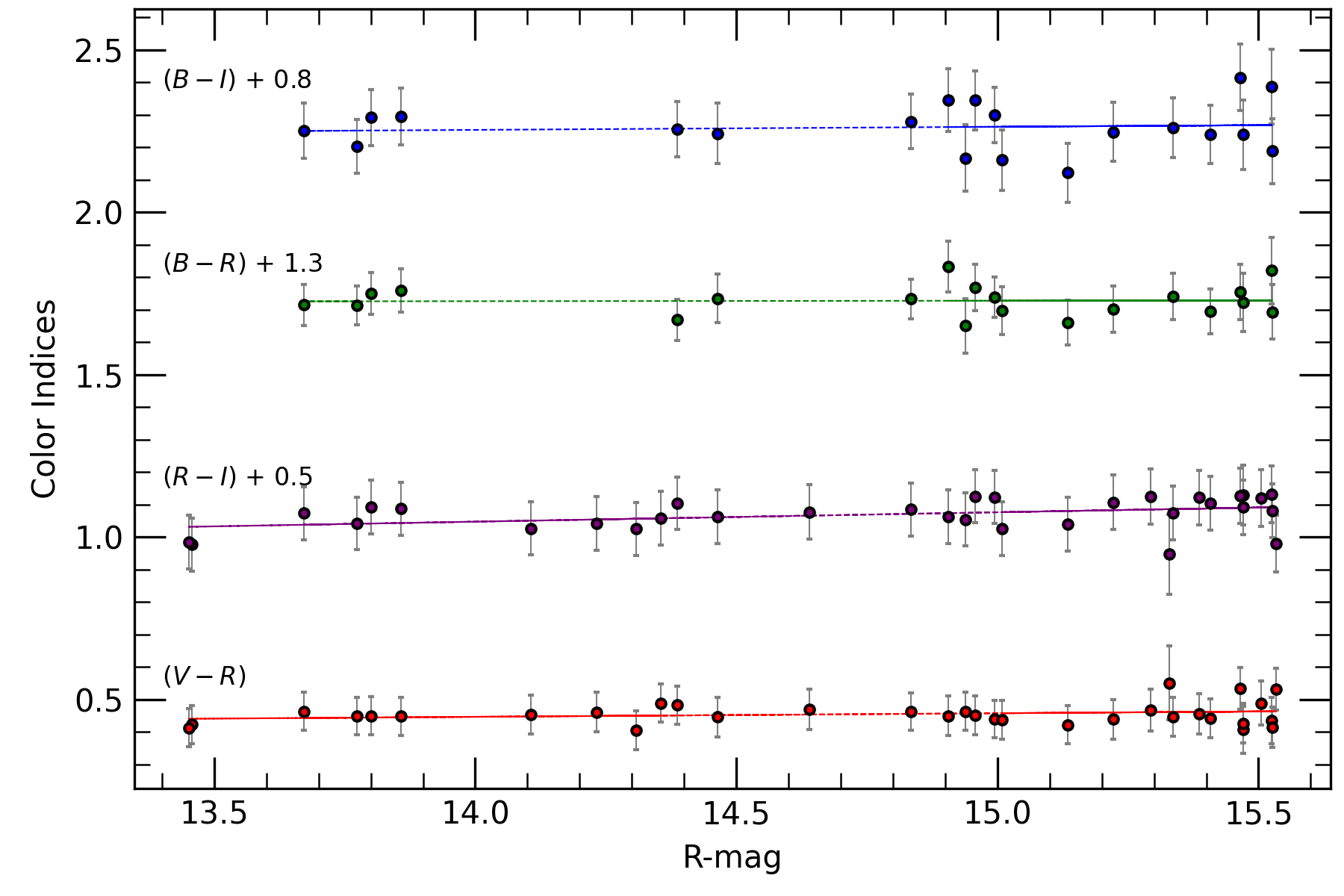}
\caption{Variation of color indices of 4C 29.45 with respect to R-magnitude.} 
\label{fig:color}
\end{figure}

In the context of blazars, the flux variation can be linked to the color variations and can also be correlated with changes in the spectral characteristics.
Thus, we
searched for a relationship in color-magnitude diagrams (CMD), which is the
distribution between the color indices (i.e. $B-I$, $B-R$, $R-I$, and $V-I$) and
the R band magnitudes (Figure \ref{fig:color}). {Since the CMDs show a linear trend, we fitted the CMDs as described in Section \ref{sec:2_5}.} The derived fit parameters are listed in Table\,\ref
{tab:color}. 
The presence of a positive relationship between two quantities is deduced from a positive slope that demonstrates a dominant BWB trend while a negative slope suggests a dominant "RWB" trend. The null hypothesis probability ($p$) may increase due to random noise, whereas a low p-value ($p \le 0.05$) implies a high likelihood of a genuine correlation. 
Additionally, the correlation coefficient reflects the significance of the connection, with $r \sim |1|$ indicating a strong correlation.  
We found that there is no significant correlation for $B-I$, $B-V$, and $V-R$ vs R-band while a weak positive correlation ($r=0.4$, $p \le 0.05$) was detected between $R-I$ and R-mag.
{However, the slope values for color indices are less than or nearly equal to their uncertainties. These results are consistent with achromatic trend as obtained from $\alpha$ vs $R$-mag.}
\citet{2010MNRAS.404.1992R} reported a significant negative correlation (RWB trend) between the optical colors and V magnitudes of the blazar 4C 29.45 during its low (faint) phase. However, our observations focus on the brighter (active) phase of the blazar. According to \citet{2010PASJ...62..645S}, the color behavior of blazars exhibits a difference based on various phases, such as outburst, active, and faint. {When the jet overcomes the accretion disk, the color variability flattens, resulting in an achromatic flux increase. This effect, known as "color saturation," has been observed in various FSRQs \citep[see][]{2006A&A...453..817V, 2022MNRAS.511.5611O}. As reported in \citep{2022MNRAS.511.5611O}, FSRQs typically exhibit more complicated color behaviors: an RWB trend during low flux states, followed by a color saturation after an achromatic brightening. In a few cases, FSRQs even display a final BWB trend during their brightest states. This is consistent with our results.} Therefore, the color/spectral behavior of 4C 29.45 can be described as achromatic trend dominant during its bright phase, and RWB during its faint phase.

\subsection{Cross-correlation analysis}
The detection of correlations between different light curves is an effective method for gaining insight into the source of emissions and the underlying radiation processes. This approach can also be utilized to determine the time lag resulting from radiation emissions from discrete spatial regions. The shortest possible variability timescale is linked to changes in the emission region of the jet. In these regions, Doppler beaming can lead to a reduction in the timescale.

\begin{figure}
\centering
\includegraphics[width=0.48\linewidth,clip=true]{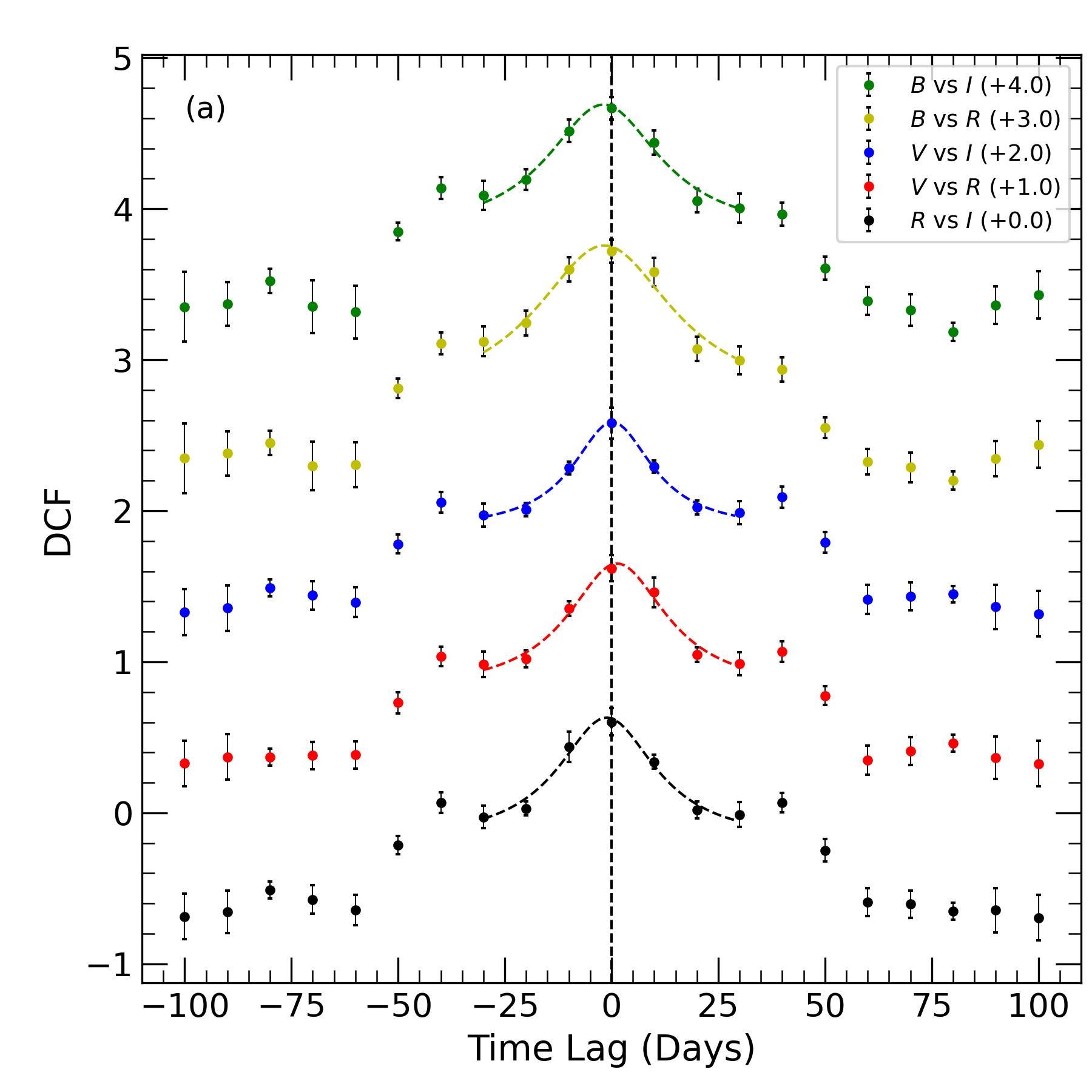}
\includegraphics[width=0.48\linewidth,clip=true]{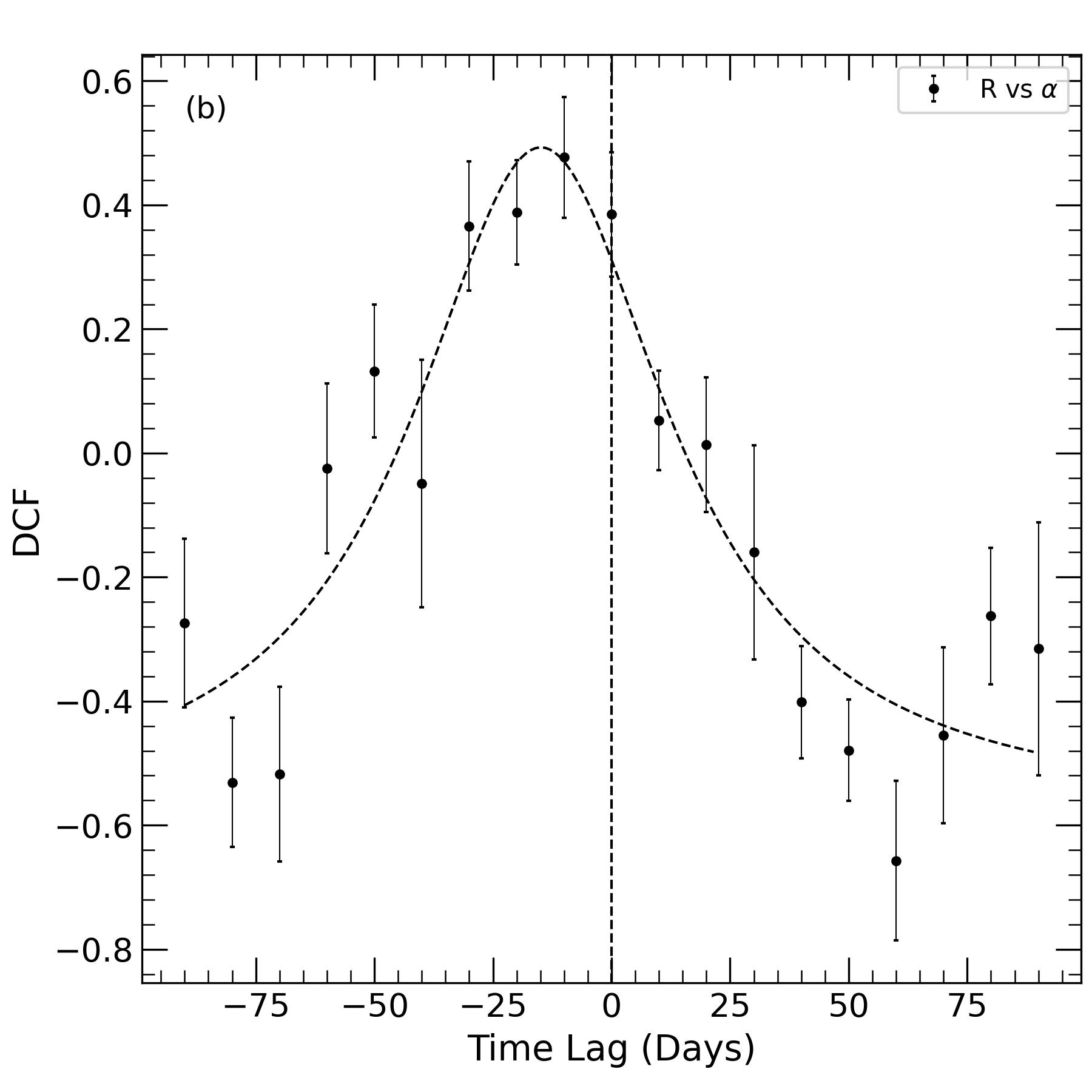}
\caption{\label{fig:dcf_fig} (a) Discrete correlation functions (DCFs) between optical light curves, (b) DCF between R-band and $\alpha$.}
\end{figure}

We used the DCF method to survey the correlation both between fluxes in optical bands, and between spectral indexes and R-band light-curve with a lag bin size of 10 days
for our observation period. The DCF distributions between optical light curves are given in
Figure \ref{fig:dcf_fig}a. Maximum DCF values of all combinations for optical
light curves range from $\sim0.6$ to $\sim0.7$ at a time lag of zero,
implying modest correlations with no time lag between optical bands. To
determine the time lag more accurately, DCF distributions where time lag
between --30 and +30 days were fitted a Gaussian profile. We found the peak
centers of the Gaussian fits as $-1.8\pm1.03$, $-1.7\pm1.5$, $0.2\pm0.6$,
$1.3\pm1.0$, and $-1.1\pm1.0$ days for the DCFs between $B$ vs $I$, $B$ vs $R$,
$V$ vs $I$, $V$ vs $R$, and $R$ vs $-I$ bands, respectively. These values
indicate that the $R$- and $I$-band variations seemingly follow B-band variations while $R$-band variations led the $V$ and $I$ variations. 
We performed the same DCF analysis for $R$-band light curve and spectral indices $\alpha$ (Figure \ref{fig:dcf_fig}b). The DCF show nonuniform distribution with larger
errors, while the peak of the DCF equals a value of $0.48\pm0.10$ at time lag
of $-10$ days. The Gaussian profile shows a centroid at a time lag of
$-14.8\pm3.06$ days, which indicates flux variability follows after spectral
variability. There are modest correlations with time gap of a few days among optical bands and between flux and spectral variability
predicated on the DCF centroid points. However, considering the sizable
uncertainty and little time lag respective to the data set with lag bin size (10 days), the obtained time lags are unreliable and might be misleading.
Since \citet{2022ApJ...926..180H} determined that variations of the $\gamma$-ray emission
lead to variations in the optical bands by $\sim -0.8 \pm 0.5$ days, the DCF analysis should
be investigated on shorter and longer timescales to ensure time lag existence
between optical bands. The time lag between optical bands is
expected to be from several hours to days instead of weeks or months if it exists.

\subsection{Periodicity search}
\begin{figure}
\centering
\includegraphics[width=0.48\linewidth,clip=true]{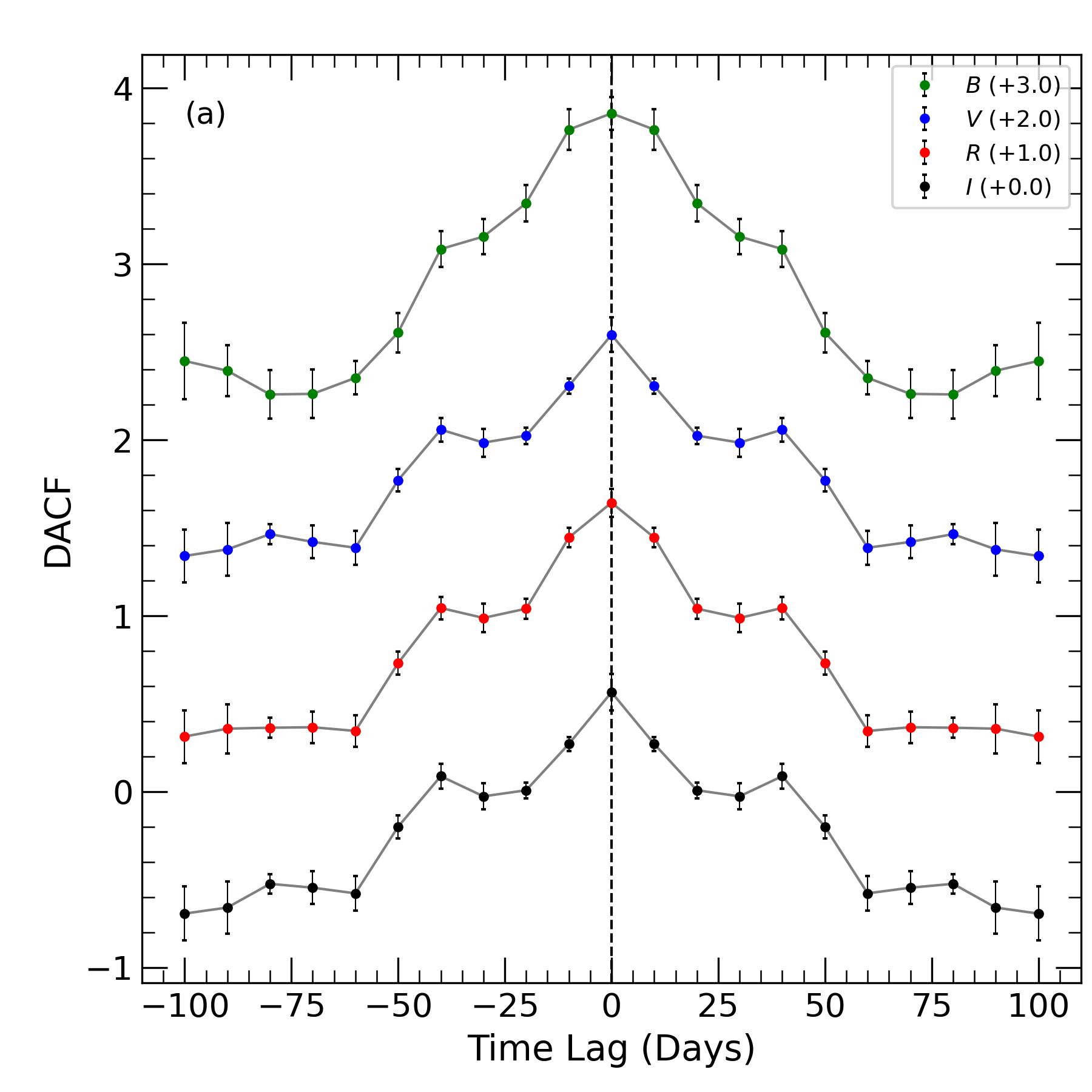}
\includegraphics[width=0.48\linewidth,clip=true]{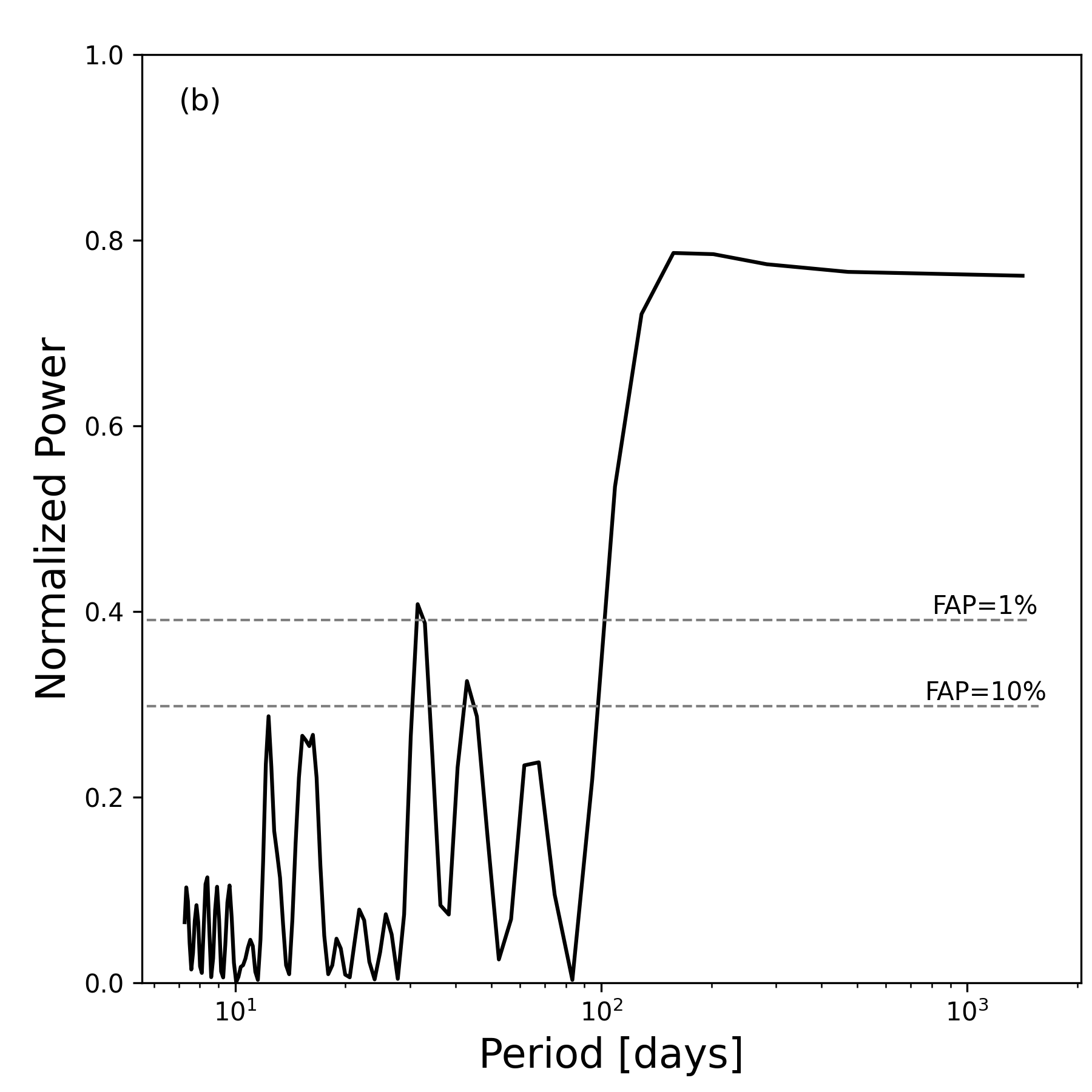}
\caption{\label{fig:dcaf_LS}  (a) Discrete autocorrelation function (DACF) analysis among optical light curves, (b) Lomb--Scargle periodogram (LSP) of 4C 29.45 for R bands.}
\end{figure}

By performing cross-correlation analyses with the DACF method on identical time series in optical bands of 4C 29.45, we aimed to detect any periodicity in the STV light curves. As the time series are identical, the DACF would show a peak at time lags equal to zero and secondary peaks indicating periodicity if present. A strong correlation is depicted by peaks with $DACF\geq0.5$. However, our results show that there were no strong secondary peaks in the DACF distributions as shown in Figure \ref{fig:dcaf_LS}a.
Additionally, we employed the LSP technique to assess the existence of periodicity in the optical R-band light curve. A significant Gaussian signal above the false alarm limit would indicate periodicity in the blazar. However, our LSP analysis (Figure \ref{fig:dcaf_LS}b) revealed no significant signals. The possibility of periodicity may be greater than $100$ days, if exists.
Therefore, to determine the periodicity of 4C 29.45, it is recommended to analyze light curves with much longer timescales (years to decades) in agreement with the reported periods of 3.55 or 1.58 years in the long-term optical light curve by \citet{2006PASJ...58..797F}.

\section{Conclusion}

In this work, we present the results of our observation period from February 2022 to July 2022, which aimed to examine the short-term variability of the blazar 4C 29.45 in optical bands. The conclusion can be summarized as follows:

\begin{itemize}
    \item {Throughout the observation period, the source was in high-phase and exhibited mild flux variability on short timescales across all BVRI optical bands considering the literature \citep{2006PASJ...58..797F,2022ApJ...926..180H}. The STV amplitude was found as $210\%$ (i.e. $\sim2.1$ mag) and the mean magnitude in R-band was 14.7 mag.
    }
    \item {A weak flare was observed to reach its peak $R=13.4$ mag on June 19th, 2022 which is fainter than the historical brightest peak \citep[i.e. R=12.4 mag,][]{2006PASJ...58..797F}. However, it should be noted that our observations do not fully cover this brightening phase of flare. The decline in the flare was approximately 0.14 magnitude per day, and its behavior was consistent across all BVRI bands.}
    \item We also investigated the variation of the optical SED from multiple data sets obtained simultaneously.  
    The results showed that nightly spectral indices varied from 1.032 to 1.573, and the mean spectral index is $1.313\pm0.070$. We found weak correlation between spectral index and R mag with a very small slope $0.0089\pm0.032$.  We also searched for correlations between the color indices and R-band magnitude. 
    {We found nonsignificant ($R\lesssim 0.5$ and $p\leq0.05$) correlations, and the obtained slopes is lesser than the fitted uncertainties. Those results indicate that the source exhibits an achromatic trend during our observations (i.e. its bright phase). However, RWB trend was found during faint phase of the source by \citet{2010MNRAS.404.1992R}.  Therefore, the jet contribution differs within the phases; the jet flux dominating the one from the disk during bright state leads to achromatic flux increase and color saturation. }
    \item The cross-correlation analysis revealed that a modest relation exists between all simultaneous optical light curves and between spectral index and R-band light curve. We found a probable time lag from hours to days with large uncertainties. If we assume a time lag exists from several hours to days in optical bands, the emission sites are not cospatial.
    \item No significant periodicity signal was found for STV, but the clue on periodicity  $\>100$ days was found consistent with reported periods of 3.55 or 1.58 years from longer timescaled light curves.
\end{itemize}

To better  understand the variability of the blazar 4C 29.45 in optical window, optical multiband studies on diverse timescales with densely sampled observations are needed. To gain a comprehensive understanding of its behavior across the entire electromagnetic spectrum, correlation analysis between the observations at various bands (i.e. radio, IR, optical, UV, X-ray, Gamma) should be performed and studied as well. Thus, it will help us draw the picture of the emission regions and constrain the theoretical models.

\section*{Acknowledgments}
We thank TÜBİTAK National Observatory for partial support in using T60
telescope with project number 22AT60-1907. AO were supported by The
Scientific and Technological Research Council of Türkiye (TÜBİTAK) through
project number 121F427.

\clearpage

\begin{table}
\centering
\caption{Observation log of the blazar 4C 29.45. The columns are (1) the date of
observations and (2) the number of data points in each filter on a particular night.
Columns 3 and 4 are the same as columns 1 and 2, respectively.}
\label{tab:obs_log} 
\begin{tabular}{cccccccccc} 
\hline\hline 
  Date of           & \multicolumn{4}{c}{Number of} & Date of     & \multicolumn{4}{c}{Number of}  \\
observations        & \multicolumn{4}{c}{data points} & observations     & \multicolumn{4}{c}{data points}  \\
(yyyy-mm-dd)     & $B$ & $V$ & $R$ & $I$ & (yyyy-mm-dd)  & $B$ & $V$ & $R$ & $I$   \\
\hline
2022-02-06 & 2 & 2 & 2 & 2 & 2022-04-23 & 1 & 1 & 1 & 1\\ 
2022-02-07 & 2 & 2 & 2 & 2 & 2022-04-24 & 1 & 1 & 1 & 1\\ 
2022-02-12 & 2 & 2 & 2 & 2 & 2022-04-25 & 1 & 1 & 1 & 1\\ 
2022-02-15 & 2 & 2 & 2 & 2 & 2022-04-27 & 0 & 1 & 1 & 1\\ 
2022-02-17 & 0 & 1 & 0 & 0 & 2022-04-28 & 1 & 1 & 1 & 1\\ 
2022-02-18 & 2 & 2 & 0 & 0 & 2022-05-07 & 1 & 1 & 1 & 1\\ 
2022-02-25 & 1 & 0 & 0 & 0 & 2022-05-08 & 1 & 1 & 1 & 1\\ 
2022-02-26 & 1 & 2 & 2 & 2 & 2022-05-28 & 0 & 1 & 1 & 1\\ 
2022-03-21 & 2 & 2 & 2 & 2 & 2022-05-29 & 1 & 1 & 1 & 1\\ 
2022-03-25 & 2 & 2 & 2 & 2 & 2022-05-30 & 1 & 1 & 1 & 1\\ 
2022-03-26 & 2 & 2 & 2 & 2 & 2022-05-31 & 1 & 1 & 1 & 1\\ 
2022-03-27 & 2 & 2 & 2 & 2 & 2022-06-01 & 1 & 1 & 1 & 1\\ 
2022-03-29 & 2 & 2 & 2 & 2 & 2022-06-18 & 0 & 1 & 1 & 1\\ 
2022-03-30 & 0 & 2 & 2 & 2 & 2022-06-19 & 0 & 1 & 1 & 1\\ 
2022-04-08 & 0 & 1 & 1 & 1 & 2022-06-21 & 0 & 1 & 1 & 1\\ 
2022-04-10 & 0 & 1 & 1 & 1 & 2022-06-22 & 0 & 1 & 1 & 1\\ 
2022-04-13 & 1 & 1 & 0 & 0 & 2022-06-23 & 0 & 1 & 1 & 1\\ 
2022-04-14 & 0 & 1 & 0 & 1 & 2022-06-26 & 0 & 1 & 1 & 1\\ 
2022-04-20 & 0 & 0 & 0 & 1 & 2022-06-27 & 0 & 1 & 1 & 1\\ 
2022-04-22 & 0 & 1 & 1 & 1 &  &  &  &  & \\ 
\hline
\end{tabular}
\\
\footnotesize{Total number of data points in the B, V, R, and I bands are 33, 49, 44, and 46, respectively.}
\end{table}

\clearpage

\begin{table}
\caption{Straight line fits to optical SEDs of the blazar 4C 29.45.} 
\label{tab:sed} 
\centering 
\resizebox{\textwidth} {!}{ 
\begin{tabular}{cccccccccc} 
\hline\hline 
Date & $\alpha$ & $C$ & $r$ & $p$  & 	Date & $\alpha$ & $C$ & $r$ & $p$ \\
\hline
2022-02-06 & $1.24\pm 0.04$ & $-6.85\pm 0.54$ & -0.999 & 8.89e-04 &	2022-04-27 & $1.38\pm 0.09$ & $-5.44\pm 1.29$ & -0.998 & 4.05e-02\\
2022-02-07 & $1.31\pm 0.04$ & $-5.74\pm 0.64$ & -0.999 & 1.11e-03 &	2022-04-28 & $1.28\pm 0.03$ & $-6.91\pm 0.43$ & -0.999 & 5.28e-04\\
2022-02-12 & $1.36\pm 0.02$ & $-5.01\pm 0.23$ & -1.000 & 1.34e-04 &	2022-05-07 & $1.17\pm 0.04$ & $-8.31\pm 0.54$ & -0.999 & 9.93e-04\\
2022-02-15 & $1.37\pm 0.01$ & $-4.99\pm 0.17$ & -1.000 & 6.87e-05 &	2022-05-08 & $1.57\pm 0.07$ & $-2.62\pm 1.07$ & -0.998 & 2.13e-03\\
2022-02-26 & $1.32\pm 0.10$ & $-5.83\pm 1.40$ & -0.995 & 5.16e-03 &	2022-05-28 & $1.29\pm 0.29$ & $-6.78\pm 4.19$ & -0.976 & 1.39e-01\\
2022-03-21 & $1.29\pm 0.02$ & $-6.33\pm 0.33$ & -1.000 & 3.10e-04 &	2022-05-29 & $1.20\pm 0.04$ & $-8.06\pm 0.63$ & -0.999 & 1.25e-03\\
2022-03-25 & $1.35\pm 0.03$ & $-5.62\pm 0.48$ & -0.999 & 5.87e-04 &	2022-05-30 & $1.11\pm 0.05$ & $-9.27\pm 0.72$ & -0.998 & 1.93e-03\\
2022-03-26 & $1.44\pm 0.02$ & $-4.34\pm 0.29$ & -1.000 & 1.86e-04 &	2022-05-31 & $1.19\pm 0.08$ & $-8.01\pm 1.20$ & -0.995 & 4.64e-03\\
2022-03-27 & $1.37\pm 0.04$ & $-5.37\pm 0.58$ & -0.999 & 8.36e-04 &	2022-06-01 & $1.44\pm 0.07$ & $-4.35\pm 1.09$ & -0.997 & 2.66e-03\\
2022-03-29 & $1.30\pm 0.06$ & $-6.56\pm 0.83$ & -0.998 & 1.89e-03 &	2022-06-18 & $1.04\pm 0.12$ & $-9.71\pm 1.69$ & -0.994 & 7.07e-02\\
2022-03-30 & $1.51\pm 0.01$ & $-3.50\pm 0.19$ & -1.000 & 5.56e-03 &	2022-06-19 & $1.03\pm 0.09$ & $-9.75\pm 1.33$ & -0.996 & 5.60e-02\\
2022-04-08 & $1.47\pm 0.00$ & $-4.07\pm 0.05$ & -1.000 & 1.57e-03 &	2022-06-21 & $1.23\pm 0.11$ & $-7.10\pm 1.55$ & -0.996 & 5.45e-02\\
2022-04-10 & $1.54\pm 0.05$ & $-3.06\pm 0.76$ & -0.999 & 2.13e-02 &	2022-06-22 & $1.39\pm 0.12$ & $-4.88\pm 1.82$ & -0.996 & 5.66e-02\\
2022-04-22 & $1.25\pm 0.36$ & $-7.32\pm 5.20$ & -0.962 & 1.76e-01 &	2022-06-23 & $1.12\pm 0.03$ & $-8.87\pm 0.46$ & -1.000 & 1.78e-02\\
2022-04-23 & $1.49\pm 0.03$ & $-3.80\pm 0.47$ & -1.000 & 4.52e-04 &	2022-06-26 & $1.28\pm 0.10$ & $-6.37\pm 1.47$ & -0.997 & 4.96e-02\\
2022-04-24 & $1.29\pm 0.06$ & $-6.78\pm 0.91$ & -0.998 & 2.28e-03 &	2022-06-27 & $1.39\pm 0.07$ & $-4.93\pm 1.04$ & -0.999 & 3.24e-02\\
2022-04-25 & $1.32\pm 0.02$ & $-6.30\pm 0.27$ & -1.000 & 1.99e-04 &	& & & & \\
\hline 
\end{tabular}
}
\\
\footnotesize{$\alpha$ = spectral index and $C$ = intercept of log($F_{\nu}$) 
against log($\nu$); $r$ = Correlation coefficient; $p$ = null hypothesis probability}
\end{table}

\clearpage

\begin{table}
\caption{Results of STV analysis of 4C 29.45.}            
\label{tab:STV_res}                   
\centering 
\resizebox{\textwidth} {!}{                     
\begin{tabular}{cccccc}           
\hline\hline                		 
Band &  Brightest state magnitude/MJD  &  Faintest state magnitude/MJD  & Average magnitude  &   Variability amplitude (\%)\\
\hline
B & - & 16.480 $\pm$ 0.089 / 59693.3271 & - & -\\ 
V & 13.864 $\pm$ 0.043 / 59750.28346 & 16.066 $\pm$ 0.046 / 59728.29339 & 15.147 $\pm$ 0.046 & 220.155\\ 
R & 13.451 $\pm$ 0.041 / 59750.28447 & 15.534 $\pm$ 0.046 / 59728.29494 & 14.705 $\pm$ 0.044 & 208.243\\ 
I & 12.967 $\pm$ 0.071 / 59750.28526 & 15.054 $\pm$ 0.074 / 59728.29597 & 14.161 $\pm$ 0.072 & 208.692\\ 
\hline                          
\end{tabular}
}
\end{table}

\clearpage

\begin{table}
\caption{Variation of color indices with R-mag.} 
\label{tab:color} 
\centering 
\begin{tabular}{ccccc} 
\hline\hline 
color indices & $m^a$ & $c^a$ & $r^a$ & $p^a$ \\
\hline 
($B-I$) & 0.96e-02 $\pm$ 2.78e-02 &  1.32 $\pm$   0.41 &   0.08 & 0.73 \\
($B-V$) & 0.14e-02 $\pm$ 1.76e-02 &  0.41 $\pm$   0.26 &   0.02 & 0.93 \\
($R-I$) & 2.89e-02 $\pm$ 1.19e-02 &  0.14 $\pm$   0.17 &   0.40 & 0.02 \\
($V-R$) & 1.10e-02 $\pm$ 0.88e-02 &  0.29 $\pm$   0.13 &   0.22 & 0.22 \\
\hline 
\end{tabular}\\
$^am_2$ = slope and $c_2$ = intercept of $\alpha$ against time or R magnitude; $r$ = Correlation coefficient; $p$ = null hypothesis probability.
\end{table}

\clearpage

\begin{thebibliography}{99}
\bibitem[\protect\citeauthoryear{Urry \& Padovani}{1995}]{1995PASP..107..803U} Urry C.~M., Padovani P., 1995, PASP, 107, 803. doi:10.1086/133630
\bibitem[\protect\citeauthoryear{Wagner \& Witzel}{1995}]{1995ARA&A..33..163W} Wagner S.~J., Witzel A., 1995, ARA\&A, 33, 163. doi:10.1146/annurev.aa.33.090195.001115
\bibitem[\protect\citeauthoryear{Valtonen et al.}{2008}]{2008Natur.452..851V} Valtonen M.~J., Lehto H.~J., Nilsson K., Heidt J., Takalo L.~O., Sillanp{\"a}{\"a} A., Villforth C. et al., 2008, Natur, 452, 851. doi:10.1038/nature06896
\bibitem[\protect\citeauthoryear{Xie et al.}{2004}]{2004MNRAS.348..831X} Xie G.~Z., Zhou S.~B., Li K.~H., Dai H., Chen L.~E., Ma L., 2004, MNRAS, 348, 831. doi:10.1111/j.1365-2966.2004.07396.x
\bibitem[\protect\citeauthoryear{Gupta et al.}{2008}]{2008AJ....135.1384G} Gupta A.~C., Fan J.~H., Bai J.~M., Wagner S.~J., 2008, AJ, 135, 1384. doi:10.1088/0004-6256/135/4/1384
\bibitem[\protect\citeauthoryear{Agarwal et al.}{2022}]{2022ApJ...933...42A} Agarwal A., Pandey A., {\"O}zd{\"o}nmez A., Ege E., Kumar Das A., Karakulak V., 2022, ApJ, 933, 42. doi:10.3847/1538-4357/ac6cef
\bibitem[\protect\citeauthoryear{Marscher \& Travis}{1996}]{1996A&AS..120C.537M} Marscher A.~P., Travis J.~P., 1996, A\&AS, 120, 537
\bibitem[\protect\citeauthoryear{Wiita}{1996}]{1996ASPC..110...42W} Wiita P.~J., 1996, ASPC, 110, 42
\bibitem[\protect\citeauthoryear{Schneider \& Weiss}{1987}]{1987A&A...171...49S} Schneider P., Weiss A., 1987, A\&A, 171, 49
\bibitem[\protect\citeauthoryear{Raiteri et al.}{2017}]{2017Natur.552..374R} Raiteri C.~M., Villata M., Acosta-Pulido J.~A., Agudo I., Arkharov A.~A., Bachev R., Baida G.~V. et al., 2017, Natur, 552, 374. doi:10.1038/nature24623
\bibitem[\protect\citeauthoryear{Ciprini et al.}{2003}]{2003A&A...400..487C} Ciprini S., Tosti G., Raiteri C.~M., Villata M., Ibrahimov M.~A., Nucciarelli G., Lanteri L., 2003, A\&A, 400, 487. doi:10.1051/0004-6361:20030045
\bibitem[\protect\citeauthoryear{Agarwal \& Gupta}{2015}]{2015MNRAS.450..541A} Agarwal A., Gupta A.~C., 2015, MNRAS, 450, 541. doi:10.1093/mnras/stv625
\bibitem[\protect\citeauthoryear{Ghisellini et al.}{1997}]{1997A&A...327...61G} Ghisellini G., Villata M., Raiteri C.~M., Bosio S., de Francesco G., Latini G., Maesano M. et al., 1997, A\&A, 327, 61
\bibitem[\protect\citeauthoryear{Villata et al.}{2002}]{2002A&A...390..407V} Villata M., Raiteri C.~M., Kurtanidze O.~M., Nikolashvili M.~G., Ibrahimov M.~A., Papadakis I.~E., Tsinganos K. et al., 2002, A\&A, 390, 407. doi:10.1051/0004-6361:20020662
\bibitem[\protect\citeauthoryear{Rani et al.}{2010}]{2010MNRAS.404.1992R} Rani B., Gupta A.~C., Strigachev A., Bachev R., Wiita P.~J., Semkov E., Ovcharov E. et al., 2010, MNRAS, 404, 1992. doi:10.1111/j.1365-2966.2010.16419.x
\bibitem[\protect\citeauthoryear{Agarwal et al.}{2019}]{2019MNRAS.488.4093A} Agarwal A., Cellone S.~A., Andruchow I., Mammana L., Singh M., Anupama G.~C., Mihov B. et al., 2019, MNRAS, 488, 4093. doi:10.1093/mnras/stz1981
\bibitem[\protect\citeauthoryear{Villata et al.}{2006}]{2006A&A...453..817V} Villata M., Raiteri C.~M., Balonek T.~J., Aller M.~F., Jorstad S.~G., Kurtanidze O.~M., Nicastro F. et al., 2006, A\&A, 453, 817. doi:10.1051/0004-6361:20064817
\bibitem[\protect\citeauthoryear{Gu \& Ai}{2011}]{2011A&A...528A..95G} Gu M.-F., Ai Y.~L., 2011, A\&A, 528, A95. doi:10.1051/0004-6361/201016280 
\bibitem[\protect\citeauthoryear{Agarwal et al.}{2021}]{2021A&A...645A.137A} Agarwal A., Mihov B., Andruchow I., Cellone S.~A., Anupama G.~C., Agrawal V., Zola S. et al., 2021, A\&A, 645, A137. doi:10.1051/0004-6361/202039301 
\bibitem[\protect\citeauthoryear{Gupta et al.}{2016}]{2016MNRAS.458.1127G} Gupta A.~C., Agarwal A., Bhagwan J., Strigachev A., Bachev R., Semkov E., Gaur H. et al., 2016, MNRAS, 458, 1127. doi:10.1093/mnras/stw377
\bibitem[\protect\citeauthoryear{Isler et al.}{2017}]{2017ApJ...844..107I} Isler J.~C., Urry C.~M., Coppi P., Bailyn C., Brady M., MacPherson E., Buxton M. et al., 2017, ApJ, 844, 107. doi:10.3847/1538-4357/aa79fc
\bibitem[\protect\citeauthoryear{Negi et al.}{2022}]{2022MNRAS.510.1791N} Negi V., Joshi R., Chand K., Chand H., Wiita P., Ho L.~C., Singh R.~S., 2022, MNRAS, 510, 1791. doi:10.1093/mnras/stab3591
\bibitem[\protect\citeauthoryear{Ghosh et al.}{2000}]{2000ApJS..127...11G} Ghosh K.~K., Ramsey B.~D., Sadun A.~C., Soundararajaperumal S., 2000, ApJS, 127, 11. doi:10.1086/313313
\bibitem[\protect\citeauthoryear{Stalin et al.}{2006}]{2006MNRAS.366.1337S} Stalin C.~S., Gopal-Krishna, Sagar R., Wiita P.~J., Mohan V., Pandey A.~K., 2006, MNRAS, 366, 1337. doi:10.1111/j.1365-2966.2005.09939.x
\bibitem[\protect\citeauthoryear{B{\"o}ttcher et al.}{2007}]{2007ApJ...670..968B} B{\"o}ttcher M., Basu S., Joshi M., Villata M., Arai A., Aryan N., Asfandiyarov I.~M. et al., 2007, ApJ, 670, 968. doi:10.1086/522583
\bibitem[\protect\citeauthoryear{Poon, Fan, \& Fu}{2009}]{2009ApJS..185..511P} Poon H., Fan J.~H., Fu J.~N., 2009, ApJS, 185, 511. doi:10.1088/0067-0049/185/2/511

\bibitem[\protect\citeauthoryear{Wills et al.}{1983}]{1983ApJ...274...62W} Wills B.~J., Pollock J.~T., Aller H.~D., Aller M.~F., Balonek T.~J., Barvainis R.~E., Binzel R.~P. et al., 1983, ApJ, 274, 62. doi:10.1086/161426 
\bibitem[\protect\citeauthoryear{Wills et al.}{1992}]{1992ApJ...398..454W} Wills B.~J., Wills D., Breger M., Antonucci R.~R.~J., Barvainis R., 1992, ApJ, 398, 454. doi:10.1086/171869
\bibitem[\protect\citeauthoryear{Fan et al.}{2006}]{2006PASJ...58..797F} Fan J.~H., Tao J., Qian B.~C., Gupta A.~C., Liu Y., Yuan Y.-H., Yang J.-H. et al., 2006, PASJ, 58, 797. doi:10.1093/pasj/58.5.797
\bibitem[\protect\citeauthoryear{Hovatta et al.}{2007}]{2007A&A...469..899H} Hovatta T., Tornikoski M., Lainela M., Lehto H.~J., Valtaoja E., Torniainen I., Aller M.~F. et al., 2007, A\&A, 469, 899. doi:10.1051/0004-6361:20077529
\bibitem[\protect\citeauthoryear{Savolainen \& Kovalev}{2008}]{2008A&A...489L..33S} Savolainen T., Kovalev Y.~Y., 2008, A\&A, 489, L33. doi:10.1051/0004-6361:200810423
\bibitem[\protect\citeauthoryear{Hallum et al.}{2022}]{2022ApJ...926..180H} Hallum M.~K., Jorstad S.~G., Larionov V.~M., Marscher A.~P., Joshi M., Weaver Z.~R., Williamson K.~E. et al., 2022, ApJ, 926, 180. doi:10.3847/1538-4357/ac4710
\bibitem[\protect\citeauthoryear{McHardy et al.}{1990}]{1990MNRAS.246..305M} McHardy I.~M., Marscher A.~P., Gear W.~K., Muxlow T., Lehto H.~J., Abraham R.~G., 1990, MNRAS, 246, 305
\bibitem[\protect\citeauthoryear{Thompson et al.}{1995}]{1995ApJS..101..259T} Thompson D.~J., Bertsch D.~L., Dingus B.~L., Esposito J.~A., Etienne A., Fichtel C.~E., Friedlander D.~P. et al., 1995, ApJS, 101, 259. doi:10.1086/192240
\bibitem[\protect\citeauthoryear{Jorstad et al.}{2017}]{2017ApJ...846...98J} Jorstad S.~G., Marscher A.~P., Morozova D.~A., Troitsky I.~S., Agudo I., Casadio C., Foord A. et al., 2017, ApJ, 846, 98. doi:10.3847/1538-4357/aa8407
\bibitem[\protect\citeauthoryear{Branly et al.}{1996}]{1996ASPC..110..170B} Branly R., Kilgard R., Sadun A., Shcherbanovsky A., Webb J., 1996, ASPC, 110, 170
\bibitem[\protect\citeauthoryear{Noble \& Miller}{1996}]{1996ASPC..110...30N} Noble J.~C., Miller H.~R., 1996, ASPC, 110, 30 
\bibitem[\protect\citeauthoryear{Ramakrishnan et al.}{2014}]{2014MNRAS.445.1636R} Ramakrishnan V., Le{\'o}n-Tavares J., Rastorgueva-Foi E.~A., Wiik K., Jorstad S.~G., Marscher A.~P., Tornikoski M. et al., 2014, MNRAS, 445, 1636. doi:10.1093/mnras/stu1873
\bibitem[\protect\citeauthoryear{Du et al.}{2022}]{2022ATel15441....1D} Du N., Hulburt M.~L., Choi H.-E.~H., Corcoran R.~M., Balonek T.~J., 2022, ATel, 15441 
\bibitem[\protect\citeauthoryear{Bradley et al.}{2022}]{2021zndo....596036B} Bradley L., Sip{\H{o}}cz B., Robitaille T., Tollerud E., Vin{\'\i}cius Z., Deil C., Barbary K. et al., 2022, zndo. doi: 10.5281/zenodo.6825092
\bibitem[\protect\citeauthoryear{Smith et al.}{1985}]{1985AJ.....90.1184S} Smith P.~S., Balonek T.~J., Heckert P.~A., Elston R., Schmidt G.~D., 1985, AJ, 90, 1184. doi:10.1086/113824
\bibitem[\protect\citeauthoryear{Heidt \& Wagner}{1996}]{1996A&A...305...42H} Heidt J., Wagner S.~J., 1996, A\&A, 305, 42
\bibitem[\protect\citeauthoryear{Edelson \& Krolik}{1988}]{1988ApJ...333..646E} Edelson R.~A., Krolik J.~H., 1988, ApJ, 333, 646. doi:10.1086/166773
\bibitem[\protect\citeauthoryear{Pandey, Gupta, \& Wiita}{2017}]{2017ApJ...841..123P} Pandey A., Gupta A.~C., Wiita P.~J., 2017, ApJ, 841, 123. doi:10.3847/1538-4357/aa705e
\bibitem[\protect\citeauthoryear{Acciari et al.}{2021}]{2021MNRAS.504.1427A} Acciari V.~A., Ansoldi S., Antonelli L.~A., Asano K., Babi{\'c} A., Banerjee B., Baquero A. et al., 2021, MNRAS, 504, 1427. doi:10.1093/mnras/staa3727
\bibitem[\protect\citeauthoryear{Villata et al.}{2004}]{2004A&A...424..497V} Villata M., Raiteri C.~M., Aller H.~D., Aller M.~F., Ter{\"a}sranta H., Koivula P., Wiren S. et al., 2004, A\&A, 424, 497. doi:10.1051/0004-6361:20040439
\bibitem[\protect\citeauthoryear{Lomb}{1976}]{1976Ap&SS..39..447L} Lomb N.~R., 1976, Ap\&SS, 39, 447. doi:10.1007/BF00648343
\bibitem[\protect\citeauthoryear{Scargle}{1982}]{1982ApJ...263..835S} Scargle J.~D., 1982, ApJ, 263, 835. doi:10.1086/160554
\bibitem[\protect\citeauthoryear{Bessell, Castelli, \& Plez}{1998}]{1998A&A...333..231B} Bessell M.~S., Castelli F., Plez B., 1998, A\&A, 333, 231 
\bibitem[\protect\citeauthoryear{Sasada et al.}{2010}]{2010PASJ...62..645S} Sasada M., Uemura M., Arai A., Fukazawa Y., Kawabata K.~S., Ohsugi T., Yamashita T. et al., 2010, PASJ, 62, 645. doi:10.1093/pasj/62.3.645
\bibitem[\protect\citeauthoryear{Otero-Santos et al.}{2022}]{2022MNRAS.511.5611O} Otero-Santos J., Acosta-Pulido J.~A., Becerra Gonz{\'a}lez J., Luashvili A., Castro Segura N., Gonz{\'a}lez-Mart{\'\i}n O., Raiteri C.~M. et al., 2022, MNRAS, 511, 5611. doi:10.1093/mnras/stac475


\end{thebibliography}
\end{document}